\documentclass[a4paper,fleqn]{cas-dc}

\usepackage{graphicx}
\usepackage[numbers]{natbib}
\usepackage{multirow}

\usepackage{subcaption}
\usepackage{amsmath}

\usepackage{float}
\usepackage{amsthm}
\usepackage{placeins}
\usepackage{algpseudocode}
\usepackage{algorithm}

\def\tsc#1{\csdef{#1}{\textsc{\lowercase{#1}}\xspace}}
\tsc{WGM}
\tsc{QE}
\tsc{EP}
\tsc{PMS}
\tsc{BEC}
\tsc{DE}
\usepackage{bm}
\usepackage{multirow}
\usepackage{multicol}

\begin{document}
\let\WriteBookmarks\relax
\def\floatpagepagefraction{1}
\def\textpagefraction{.001}

%
%
%

%%% Title part
\shorttitle{SEAGET: Seasonal and Active hours guided Graph Enhanced Transformer for the next POI recommendation}
\shortauthors{A.A. Hasan et~al.}

\title [mode = title]{SEAGET: Seasonal and Active hours guided Graph Enhanced Transformer for the next POI recommendation}                      

%
%
%

%%% Author part
\author[1]{Alif Al Hasan} [orcid=0009-0000-3752-616X] \cormark[1]
\ead{alif.stu2017@juniv.edu}

\author[1]{Md. Musfique Anwar}[orcid=0000-0001-5159-1865]
\ead{manwar@juniv.edu}

\address[1]{Department of Computer Science and Engineering, Jahangirnagar University, Savar, Dhaka }

\cortext[cor1]{Corresponding author}
\iffalse
\cortext[cor2]{Principal corresponding author}
\fntext[fn1]{This is the first author footnote.}
\fntext[fn2]{Another author footnote.}

\nonumnote{This note has no numbers.}
\fi

%
%
%

%%% Abstract part
\begin{abstract}
One of the most important challenges for improving personalized services in industries like tourism is predicting users' near-future movements based on prior behavior and current circumstances. Next POI (Point of Interest) recommendation is essential for helping users and service providers by providing personalized recommendations. The intricacy of this work, however, stems from the requirement to take into consideration several variables at once, such as user preferences, time contexts, and geographic locations. POI selection is also greatly influenced by elements like a POI's operational status during desired visit times, desirability for visiting during particular seasons, and its dynamic popularity over time. POI popularity is mostly determined by check-in frequency in recent studies, ignoring visitor volumes, operational constraints, and temporal dynamics. These restrictions result in recommendations that are less than ideal and do not take into account actual circumstances. We propose the Seasonal and Active hours-guided Graph-Enhanced Transformer (SEAGET) model as a solution to these problems. By integrating variations in the seasons, operational status, and temporal dynamics into a graph-enhanced transformer framework, SEAGET capitalizes on redefined POI popularity. This invention gives more accurate and context-aware next POI predictions, with potential applications for optimizing tourist experiences and enhancing location-based services in the tourism industry.
\end{abstract}

\maketitle

%
%
%

%%% Introduction part
\section{Introduction}

Location-based social networks (LBSNs), have seen a lot of development in the last several years. Examples of these include Yelp and Foursquare. Users have the option to share their locations and experiences with friends by checking in at sites that they find interesting. A check-in record normally contains the visited point of interest (POI) as well as any supplementary contexts (such as categories, timestamps, and GPS) that characterize the user's movement. The massive volume of check-in data generated by millions of users in large-scale social networks such as in LBSNs is an excellent opportunity to explore the underlying trends in user check-in behavior. This massive amount of data is utilized by numerous point-of-interest and trip recommendation systems \cite{Feng2015Personalized}. With reference to users' past and present footprints (check-ins), these systems try to forecast the next point of interest (POI) that they will visit. They assist consumers in more effectively exploring their surroundings and help businesses enhance their advertising methods \cite{Jiang2015Author}.

Almost all travel and point of interest (POI) recommender systems use popularity as one of its metrics for selecting the next POI. Most of them define a POI's popularity indicator as the quantity of check-ins. The wisest course of action isn't always to gauge popularity solely based on check-in statistics. We can make a few improvements based on this. A point's popularity fluctuates throughout time from the start. For example, let us consider two POIs namely A and B where the number of check-ins in both POIs are almost equal. In point-A, the majority of check-ins were made recently, whereas in point-B, the majority of check-ins took place in some years back. Given this scenario, point-B was presumably more popular back then, but selecting point-A over point-B makes far more sense at the moment. Secondly, not everyone will find a location appealing just because it is well-liked by a particular population. In contrast to the large number of people who visit point A on a weekly or other regular basis, just a small number of people gather at point B, but they do so daily. Because point-A is chosen by a larger spectrum of people, it is therefore possible that a newcomer would choose it over point-B, even though there are exactly the same number of check-ins at these two places. In light of these insights, we shall redefine popularity for our next POI selection problem.

The seasonal dynamics that affect the decision of which point of interest (POI) to visit next are often ignored in the studies that are already available. The way that individuals choose points of interest (POIs) is greatly influenced by seasonal and weather variations. People tend to stay away from busy beaches in the summer because of the extreme heat, while people visit forests less often in the fall because of the leaves falling off the trees diminishing their appeal. Taking the changing seasons into account when determining the likelihood of POI selection, we close this gap in our research.

Not to mention, one frequently disregarded factor in the literature that exists in the field of location-based recommendation systems is the Points of Interest's (POI) operating hours, which have a big impact on user experience. Imagine a situation in which a user asks for a recommendation for a lunch spot, yet the model only recommends a well-known restaurant that serves dinner. Similarly, offering the choice of the recommended park as an alternative in the event that it is closed on the recommended day would be counterproductive and could even aggravate the user, detracting from the intended feeling of refreshment. In order to address this issue, we have included an advanced filtering system in the last layer of our model. This enhanced function carefully assesses the hours that Points of Interest (POIs) are open, so that suggestions are in sync with users' schedules. Our goal is to improve the relevance and usefulness of our suggestions by incorporating this enhanced feature, which will also help to create a more satisfying and harmonious user experience in general.

    \subsection{Research Objectives}

    In this paper, we present SEAGET (Seasonal and Active hours guided Graph-Enhanced Transformer), a comprehensive model that extends our previous work \cite{hasan2024redefiningpopularity}, which redefined popularity measures and integrates new parameters like operational hour-based filtering and seasonal effects on POI selection, thereby improving the prediction of users' next Points of Interest (POIs). Through self-attention mechanisms, SEAGET, in contrast to conventional RNN and LSTM techniques, makes use of the transformer architecture to enable direct learning of each check-in's contribution from the input trajectory to the final recommendation. With this feature, the model may perform better by determining the importance of each individual check-in while also combining all check-ins inside the trajectory for prediction. As far as we are aware, SEAGET is a groundbreaking project because no other research has included this wide range of variables into a single POI recommendation model. 
    
    Some of the research's noteworthy contributions include the following:
    
        \begin{itemize}
            \item We have redefined popularity for the next POI or trip recommendation system.
            \item We have effectively incorporated the seasonal dynamics into our procedure for choosing which POI to visit next.
            \item We have incorporated an improved filtering process that filters out POIs according to their operating hours to improve the accuracy of our recommendations.
            \item We ran experiments using real-world dataset to demonstrate the efficacy of the proposed model.
        \end{itemize}

    \subsection{Paper Outline}
    
    In Section 2, we reviewed several studies that are relevant to our work. We made an effort to summarize their approach and set it apart from ours. In-depth analysis of the subtleties of problem formulation is provided in Section 3. All of the details of our suggested \textbf{SEAGET} framework and initial concepts are covered in Section 4. Our experimental setup was discussed in Section 5 and findings are then revealed in the next Section 6. The specific description of our findings and a summary of our overall contributions are included in Section 7. Section 8 wraps up the work and outlines our next research goals.

%
%
%

%%% Related work part
\section{Related Work}

    \subsection{Recent Advances in Recommendation Systems}
    Recent research has looked into a number of ways to make recommendation systems better. By employing a variational information bottleneck, \cite{yang2024self} suggests a self-explainable POI recommendation architecture that improves accuracy and transparency. MEGAN \cite{wang2023multi} integrates user behavior and heterogeneous substance to improve session-based recommendations by utilizing a Multi-view Enhanced Graph Attention Network. To effectively balance long- and short-term user interests, SLS-REC \cite{fu2024contrastive} integrates contrastive learning, a GNN-based geographic imbalance model, and spatio-temporal Hawkes attention. Data sparsity in session-based recommendation is addressed by CSGNN \cite{wang2024category} using self-supervised learning and a category-aware heterogeneous hypergraph. Graph differential equations and interval-aware attention are introduced by POIGDE \cite{yang2024siamese} to model dynamic user interests.

    \subsection{Next POI Recommendation}
    Next POI recommendation systems (like \cite{Cheng2013Where}) prioritize the temporal aspect of recent trajectories to predict a user's upcoming actions, in contrast to standard POI recommendation approaches. Early studies used methods like Markov chains, which are frequently used in other sequential recommendation tasks \cite{Cheng2013Where, Jihang2013What, Zhang2014LORE}. An innovative method utilizing matrix factorization with customized Markov chains (FPMC) \cite{Rendle2010Factor} was presented by Cheng et al. \cite{Cheng2013Where}. An additive Markov chain model was also presented by Zhang et al. \cite{Zhang2014LORE} in order to account for consecutive transitive influences. In addition, studies looked into modifying popular matrix factorization or metric embedding methods for the next POI suggestion \cite{Jiang2015Author, Liu2016Unified, Zhao2016STELLAR}. But in terms of handling sequence data, these earlier techniques fall short of deep neural network models.
    
    Deep learning and advanced embedding approaches have led to recent breakthroughs in POI recommendation \cite{Feng2020HME}. Several RNN variations have been developed to capture sequential correlations and temporal dynamics \cite{Liu2016Pred, Jiang2023Model, Wu2019Long, Wu2022PLong, Zhao2020Discover, Zhao2019Where}. The integration of spatial-temporal contexts into RNN layers was first demonstrated by Liu et al. in 2016 \cite{Liu2016Pred}. Geographic distance transition matrices were used to express spatial contexts, and time transition matrices were used to capture temporal context. Other applications of LSTM include modeling users' long- and short-term preferences; these examples are shown in PLSPL \cite{Wu2022PLong} and LSPL \cite{Wu2019Long}, where typical LSTM models were trained for short-term trajectory mining. To simulate time and distance intervals in both short- and long-term sequences, Zhao et al. presented STGN, a novel LSTM unit with two time gates and two distance gates \cite{Zhao2019Where}. These methods approach the issue of recommending the next POI as a sequential prediction. Studies such as DeepMove \cite{Feng2018DeepMove} and STAN \cite{Luo2021STAN} have included the attention mechanism into this task. Recurrent neural networks were utilized by DeepMove to catch sequential transitions in suggestions and to offer an attention model to capture multi-level periodicity patterns. For non-adjacent point-to-point interactions, where sequential models fail, STAN leverages the self-attention mechanism. The advantages of utilizing generic user movements have been disregarded in these investigations, nevertheless.

    \subsection{Graphs in Next POI Recommendation}

    For traditional recommendation tasks, graph-based techniques, such as those that make use of location-based social networks (LBSNs), provide a strong framework. Yuan et al. \cite{Yuan2014Graph}, for example, created the Geographical-Temporal Influences Aware Graph (GTAG), which consists of user, session, and POI nodes. Adding session nodes, however, may cause the graph's size to increase. To capture sequential impacts, geographical influences, temporal dynamics, and semantic aspects, respectively, Xie et al. \cite{Xie2016Learn} created four bipartite graphs: POI-POI, POI-Region, POI-Time, and POI-Word. They trained graph embeddings using statistical techniques and conditional probability, extending the LINE network embedding model \cite{Tang2015LINE} to bipartite graphs. Comparably, using POI-User and POI-POI graphs, Chang et al. \cite{Chang2020Learn} presented the Graph-based Geographical Latent Representation model (GGLR). Connections between user and POI nodes in the POI-User bipartite network represent user preferences. By randomly selecting the prior and subsequent check-ins from different sequences, the authors of a recent study \cite{Yang2022Discovering} included local transitions of POIs. Without specifically representing multi-hop patterns, this approach seeks to capture one-hop transitions. Furthermore, Yang et al. \cite{Yang2022GETNext} applied graph-based methods for next point-of-interest recommendation, using a consistent graph structure to capture global trends across all points of interest. In the predicted trajectory flow map, their method encoded general transitional information about POIs.

    \subsection{Transformer-based Next POI Recommendation}

    TLR-M \cite{halder2021transformer}, a multi-task transformer model with multi-head attention for queuing time-aware next POI recommendation, is one recent development in POI recommendation that incorporates long-term dependencies for enhanced performance. This method is extended by TLR-M\_UI \cite{halder2022poi}, which concurrently predicts POI suggestions and minimizes queue times by taking user interests and queuing time into account. In order to improve the accuracy of next POI prediction, GETNext \cite{Yang2022GETNext} presents a Graph Enhanced Transformer model that makes use of a global trajectory flow map and time-aware category embeddings. By combining a Transformer network with a feature-based POI grouping technique, \cite{he2023feature} improves recommendation accuracy and computational efficiency compared to graph-based models.

    \subsection{Seasonal Influence on the Next POI Recommendation}

    By focusing on annual seasonality and local-level patterns, \cite{Elena2020LSSLTT} presented a novel approach for time-aware recommendation in location-based social networks. Using real-world data, their analysis revealed significant performance increases with locality-specific seasonality, which benefited active users and areas with different seasonal weather patterns in particular. Similar to this, \cite{Trattner2018Investigate} looked into how Point of Interest (POI) suggestions in location-based social networks were affected by different seasons and weather. They demonstrated significant gains in recommendation accuracy over conventional techniques by adding weather-related parameters to the Rank-GeoFM algorithm, such as temperature, cloud cover, humidity, and intensity of precipitation.

We highlight the complex relationships between user check-ins and the recentness of those records, which contribute to the popularity of a Point of Interest (POI). We present an improved definition of popularity, in contrast to prior approaches that prioritized check-in number over user count. With a focus on recent trends over historical achievements, this definition takes into account both the overall number of users checking in and the frequency of check-ins. In addition, we examine seasonal differences in visitation patterns, providing insight into the year-round evolution of preferences. We present a filtering method based on POI operational timeframes to improve the precision of the next POI recommendations. In addition to enhancing our knowledge of POI recommendation dynamics, this thorough approach offers practical guidance for recommendation system optimization.

%
%
%

%%% Problem formulation part
\section{Problem Formulation}

Let's now investigate the major concepts the paper presents. Using the idea of users as a starting point, denoted by \(U\), we construct a collection \(U = \{u_1, u_2, \ldots, u_M\}\), where \(M\) is the total number of users. Next up are points of interest (POIs), which are various places like restaurants, hotels, coffee shops, parks, shopping centers, apparel stores, bus terminals, airports, etc. With \(N\) signifying the number of distinct POIs, they are denoted by the notation \(P = \{p_1, p_2, \ldots, p_N\}\). Within the set \(T = \{t_1, t_2, \ldots, t_K\}\), where \(K\) denotes the number of timestamps, and \(t_x\) indicates a discrete point in time.

A tuple \(p = \langle \text{cat, freq, lat and lon} \rangle\) that includes the category (cat), the frequency of visits (freq), and the geographic coordinates (latitude (lat) and longitude (lon)) specifies every POI \(p\) in set \(P\). In this instance, the category (represented by \(\text{cat}\) corresponds to predefined categories like "restaurant" or "train station."

\textbf{Definition 3.1 (Check-in)}: A check-in indicates that user \(u\) visited POI \(p\) at timestamp \(t\). It is represented by a tuple \(q = \langle u, p, t \rangle \in U \times P \times T\).

Each \(q_u^i\) represents the \(i\)-th check-in record, and the sequence of a user's check-in actions is an ordered collection of these events, represented as \(Q_u = (q_u^1, q_u^2, q_u^3, \ldots)\).

\textbf{Definition 3.2 (Check-in Set)}: The sequences of check-ins for every user that are part of the collection/set are represented as \(Q_U = \{Q_{u_1}, Q_{u_2}, \ldots, Q_{u_M}\}\).

During data preparation, we split each user's check-in sequence \(Q_u\) into successive trajectories, which are represented as \(Q_u = S_u^1 \oplus S_u^2 \oplus \ldots\), concatenating them with \(\oplus\). The varying lengths of these trajectories correspond to a series of check-ins that happen across predefined times, such as a full day.

Our primary goal is to leverage the user's current trajectory and historical check-in history to anticipate when they will visit points of interest (POIs). Our objective is to ascertain the most likely future POIs (\(q_{m+1}, q_{m+2}, \ldots, q_{m+k}\)) to be visited by \(u_i\), where \(k \geq 1\) is typically set to 1. For a given user \(u_i \in U\), given a set of historical trajectories \(\{S_u^i\}_{i \in \mathbb{N}, u \in U}\) and a current trajectory \(S' = (q_1, q_2, \ldots, q_m)\).

%
%
%

%%% SEAGET part

\begin{figure*}
    \centering
    \includegraphics[width=.9\linewidth]{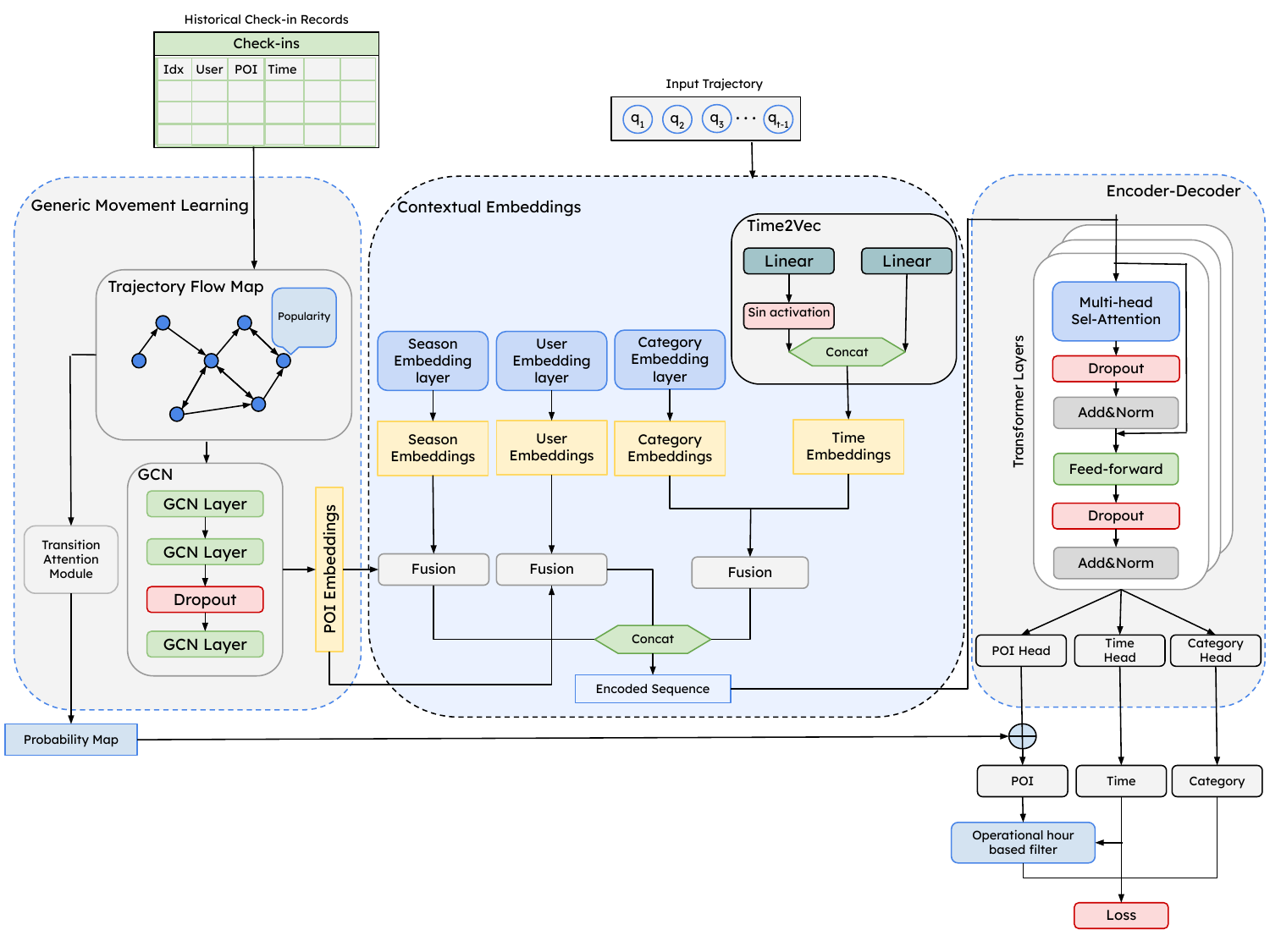}
    \caption{An overview of the SEAGET model.}
    \label{fig:seaget-model}
\end{figure*}

\section{Our proposed model: SEAGET}

We'll talk about the SEAGET technique and some related concepts in this part. 

    \subsection{Our Popularity definition}

    We define popularity more precisely by carefully balancing the number of check-ins with the diversity of users who add to these entries. In addition, we carefully balance the importance of recent check-in behavior relative to historical values, creating a sophisticated framework that can accommodate changing user preferences. Look, this is our new definition of popularity:
    
    \begin{equation}
        \label{eq:pop}
        \begin{split}
            \text{Popularity = } & \beta \left( \alpha \cdot C_{\text{user}}^{\text{rec}} + (1 - \alpha) \cdot C_{\text{chkin}}^{\text{rec}} \right) \\
            & + (1 - \beta) \left( \alpha \cdot C_{\text{user}}^{\text{past}} + (1 - \alpha) \cdot C_{\text{chkin}}^{\text{past}} \right)
        \end{split}
    \end{equation}
    
    The formula denotes the total number of unique users who checked in lately as \(C_{\text{chkin}}^{\text{rec}}\) and the number of recent user check-ins as \(C_{\text{user}}^{\text{rec}}\). However, \(C_{\text{user}}^{\text{past}}\) and \(C_{\text{chkin}}^{\text{past}}\) represent the total number of check-ins before the most recent ones and the number of unique users who checked in, respectively. The values \(\alpha\) and \(\beta\) determine the weighting factors, which signify the relative relevance of user count in relation to check-in count and recent records in relation to older ones, respectively. Real-number constraints are guaranteed for the parameters \(\alpha\) and \(\beta\), since they are both limited to the interval \textbf{\(0 \leq \alpha, \beta \leq 1\)}.

    \subsection{SEAGET Structure Overview}

    Our SEAGET model~\ref{fig:seaget-model} is influenced by the inventive structure of GETNext \cite{Yang2022GETNext}, integrating its concepts to improve the accuracy and efficacy of recommendations. By skillfully incorporating crucial components from GETNext, our model strives to attain exceptional performance in predicting user behavior and preferences.
    
    Based on historical trajectories explained in the trajectory flow map (Sect.~\ref{subsec:traj-flow-map}), the SEAGET model, which is illustrated in Figure~\ref{fig:seaget-model}, incorporates a number of critical elements within its architecture. Two significant aspects of the trajectory flow map significantly impact recommendations:

    \begin{enumerate}
        \item The utilization of the trajectory flow map is fundamental in the training process of a graph neural network (GNN), which is essential for creating embeddings of Points of Interest (POI). These embeddings include fundamental characteristics such as category, geographic location, and POI popularity, which are crucial for representing generic user movement patterns across many points of interest. The GNN analyzes the trajectory flow map to identify geographical and temporal correlations, which improves its understanding and predicting ability regarding user preferences and behavior.
        \item An attention module enhances GNN training by increasing the comprehension of user movement patterns. The resulting transition attention map effectively represents the likelihood of transitions between various points of interest (POIs), accurately reflecting the changing patterns of user interactions across time. By utilizing the adjacency matrix derived from the trajectory flow map and considering node properties, this attention mechanism provides informative probabilities for future visits to Points of Interest (POIs), enhancing the study of user behavior in location-based services.
    \end{enumerate}
    
    In addition, our system incorporates contextual modules that are crucial for understanding and interpreting important encodings, hence enhancing predictive accuracy. These modules consist of temporal encodings, POI category embeddings, seasonal embeddings and user embeddings, all intricately connected to improve the model's predictive ability (explained in Section~\ref{subsec:contex-embed}). The primary features of embedding modules encompass:
    
    \begin{itemize}
        \item The combination of both point of interest (POI) and user embeddings allows for customized recommendations that take into account the specific preferences and inclinations of each user. This approach ensures a comprehensive grasp of user preferences and delivers personalized suggestions that are aligned with their personal preferences.
        \item We recognize the importance of temporal dynamics and investigate the complex relationship between POI category embeddings and time encodings, which are essential for understanding temporal patterns in user preferences. The combination of temporal signals and categorical embeddings improves the accuracy of predictions, allowing for a better understanding of behavioral patterns such as visits to certain points of interest during peak hours.
        \item Our recommendation method effectively includes the subtle seasonal variations that influence visiting trends at various points of interest (POIs) by integrating POI embeddings with seasonal embeddings. This advanced methodology guarantees the meticulous inclusion of seasonal influences, allowing for precise predictions of the next point of interest that are specifically matched to the present season. By combining these embeddings, our system is able to understand and adjust to the changing user preferences that are influenced by seasonal variations. As a result, it improves the accuracy and relevancy of its recommendations in different time periods.
    \end{itemize}
    
    The process of creating check-in embedding vectors combines user data, POI categories, timestamps, and unique specifications to comprehensively represent trajectories. Subsequently, accurate predictions of points of interest (POI) are made by utilizing a combination of multilayer perceptron (MLP) heads and a transformer encoder. These predictions are then improved by incorporating the transition attention map through a residual link. In order to guarantee the effectiveness and applicability of suggestions, a sophisticated filtering technique is employed to carefully remove Points of Interest (POIs) that are not open during the recommended period.

    \subsection{Decoding the Trajectory Flow Map}\label{subsec:traj-flow-map}
    
    Exploring a complex methodology entails examining the details of a weighted directed graph with attributes, represented as $G = (V, E, l, w)$. This graph serves as a visual representation of the flow of trajectories. It is created using a large collection of historical trajectories, denoted as $S = \{S_u^i\}_{i \in \mathbb{N}, u \in U}$, where:
        \begin{itemize}
            \item Points of Interest (POIs) are formed by the collection of nodes $V$.
            \item Each point of interest (POI) represented by $p$ in the set $P$ contains characteristics that are enclosed within $l(p)$. These features consist of coordinates $(\text{lat, lon})$, category represented by $\text{category}$, and the frequency $\text{eq}$ of occurrence within trajectories in $S$.
            \item The edges linking $p_1$ and $p_2$ in trajectory $S_u^i$, denoted as $(p_1, p_2)$, indicate consecutive visits to Points of Interest (POI).
            \item The weight $w(p_1, p_2)$ is allocated to each edge $(p_1, p_2)$ based on their occurrences across trajectories within $S$, as defined in Equation \ref{eq:pop}.
        \end{itemize}

        \textbf{Definition (Trajectory Flow Map)}: The trajectory flow map $G$ is a directed graph with attributes and weights, denoted as $G = (V, E, l, w)$, where:
        \begin{itemize}
            \item $V$ denotes the collection of nodes that correspond to Points of Interest (POIs).
            \item $E$ represents the edges that connect Points of Interest (POIs) based on their sequential appearance in historical trajectories.
            \item $l(p)$ represents the features linked to each point of interest $p$, such as its geographical coordinates, category, and frequency of recurrence.
            \item The function $w(p_1, p_2)$ represents the weight assigned to the edge that connects points of interest (POIs) $p_1$ and $p_2$.
        \end{itemize}
    
        \subsubsection{Learning POI Embedding}
        Given the trajectory flow map $\mathcal{G}$, we aim to learn a vectorized representation of POIs that captures common POI transition patterns and attributes. To achieve this, we utilize a Graph Convolutional Network (GCN) based on the spectral GCN \cite{thomas2017GCN} approach. Starting with the adjacency matrix $\mathbf{A} \in \mathbb{R}^{N \times N}$ of $\mathcal{G}$, we compute the normalized Laplacian matrix as: 

        \begin{equation}
        \tilde{\mathbf{L}} = (\mathbf{D} + \mathbf{I}_N)^{-1} (\mathbf{A} + \mathbf{I}_N),
        \end{equation}
        
        where $\mathbf{D}$ is the degree matrix and $\mathbf{I}_N$ is the identity matrix. The input node feature matrix $\mathbf{H}^{(0)} = \mathbf{X} \in \mathbb{R}^{N \times C}$ propagates through the GCN layers according to the rule:
        
        \begin{equation}
        \mathbf{H}^{(l)} = \sigma \left( \tilde{\mathbf{L}} \mathbf{H}^{(l-1)} \mathbf{W}^{(l)} + \mathbf{b}^{(l)} \right),
        \end{equation}
        
        where $\mathbf{H}^{(l-1)}$ represents the input signals at the $(l-1)$-th layer, $\mathbf{W}^{(l)} \in \mathbb{R}^{C \times \Omega}$ is the weight matrix, $\mathbf{b}^{(l)} \in \mathbb{R}^\Omega$ is the bias vector, and $\sigma$ denotes the activation function, specifically a Leaky ReLU with a leaky rate of 0.2. To enhance expressiveness, we stack $l^*$ GCN layers and apply dropout before the final layer.
        
        The GCN module's output is computed as:
        \begin{equation}
        \mathbf{e}_p = \tilde{\mathbf{L}} \mathbf{H}^{(l^*)} \mathbf{W}^{(l^*+1)} + \mathbf{b}^{(l^*+1)} \in \mathbb{R}^{N \times \Omega},
        \end{equation}
        
        where $\mathbf{e}_p$ represents the embedding of POI $p$. These embeddings encapsulate the historical trajectory patterns and general movement trends of all users, providing rich information for downstream tasks, such as modeling users' visiting behaviors. Notably, even with short trajectories, these embeddings offer substantial predictive value.
    
        \subsubsection{Deciphering Transition Attention Map}
        To explicitly model the transition probabilities between points of interest (POIs), a transition attention map is proposed. This mechanism adjusts the final prediction by amplifying the impact of collective signals.

        Given input node features $X \in \mathbb{R}^{N \times h}$ and the graph $\mathcal{G}$, the attention map $\Phi \in \mathbb{R}^{N \times N}$ is computed as follows:
        \begin{align}
            \Phi_1 &= (X \times W_1) \times a_1 \in \mathbb{R}^{N \times 1}, \\
            \Phi_2 &= (X \times W_2) \times a_2 \in \mathbb{R}^{N \times 1}, \\
            \Phi &= (\Phi_1 \times \mathbf{1}^\top + \mathbf{1} \times \Phi_2^\top) \odot (\tilde{L} + J_N),
        \end{align}
        where $W_1, W_2 \in \mathbb{R}^{h \times h}$ are trainable feature transformation matrices, $a_1, a_2 \in \mathbb{R}^h$ are learnable vectors, $\mathbf{1} \in \mathbb{R}^{N \times 1}$ is an all-ones vector, $J_N$ is a matrix of ones, $\tilde{L}$ is the normalized Laplacian matrix shifted to range $[1, 2]$, and $\odot$ represents element-wise multiplication.
        
        The $i$-th row of $\Phi$ provides the (unnormalized) probability distribution of transitioning from the $i$-th POI to all other POIs. These transition probabilities are used to adjust recommendation results from a later transformer module.

    \subsection{Contextual Embedding Module}\label{subsec:contex-embed}
    Next point-of-interest (POI) recommendations that are tailored to individual users are created using the Contextual Embedding Module. This module blends spatiotemporal contexts with user preferences to generate personalized recommendations. This module has three crucial fusion components:
    
        \subsubsection{POI-User Embeddings Fusion}
            In order to capture both the general patterns found in POIs and the user-specific behaviors encountered in previous check-in sequences, it is essential to combine user embeddings with POIs. Using a function $\text{fembed}(u)$, we first retrieve the user embedding $e_u$ in this case, which is represented as:
            \begin{equation}
                e_u = \text{fembed}(u) \in \mathbb{R}^\Omega. \label{eq:user_embedding}
            \end{equation}
            This embedding captures user preferences and actions that are sophisticated.
            
            The process involves combining the point of interest (POI) embedding $e_p$ with the user embedding $e_u$ to create the fused embedding $e_{p,u}$. This fused embedding is subsequently utilized in the following manner:
            \begin{equation}
                e_{p,u} = \sigma(w_{p,u} [e_p ; e_u] + b_{p,u}) \in \mathbb{R}^{\Omega \times 2}, \label{eq:fused_embedding}
            \end{equation}
            The activation function is represented by the symbol $\sigma$, and the weights and bias are respectively written as $w_{p,u}$ and $b_{p,u}$. In order to enhance the model's ability to gather personalized recommendations, the concatenated vector $[e_p ; e_u]$ merges user attributes with points of interest.
            
        \subsubsection{Time-Category Embeddings Fusion}
            This fusion approach combines categorical embeddings of Points of Interest (POIs) with temporal information acquired by Time2vector. Time2vector effectively encodes time values, considering the temporal component of user behavior. Furthermore, a simultaneous utilization of an embedding layer is employed for POI categories.
            
            The fusion equation for time-category embeddings, $e_{c,t}$, is as follows:
            \begin{equation}
                e_{c,t} = \sigma(w_{c,t} [e_t ; e_c] + b_{c,t}) \in \mathbb{R}^{\Psi \times 2}, \label{eq:time_category_fusion}
            \end{equation}
            The learnable weight vector is denoted as $w_{c,t}$, while the bias is represented as $b_{c,t}$. Integrating temporal and categorical data becomes simpler when $e_t$ and $e_c$ are combined.
            
        \subsubsection{Season-POI Embeddings Fusion}
            This fusion method combines seasonal embeddings and POI embeddings to consider the impact of seasonal factors on patterns in POI selection. By accurately accounting for seasonal variations, the model is able to forecast the next point of interest (POI) based on the present season. The fusion method merges the embeddings of seasons with those of points of interest (POIs) to offer a holistic contextual understanding for the purpose of making recommendations. The fusion equation for season-POI embeddings, denoted as $e_{s,p}$, is expressed as:
            \begin{equation}
                e_{s,p} = \sigma(w_{s,p} [e_s ; e_p] + b_{s,p}) \in \mathbb{R}^{\Phi \times 2}, \label{eq:season_poi_fusion}
            \end{equation}
            The variable $w_{s,p}$ represents the weight vector that can be adjusted during the learning process, whereas $b_{s,p}$ indicates the bias. This integrated embedding merges seasonal and POI (Point of Interest) data to improve the model's capacity to capture fluctuations in user preferences and behavior that occur over different seasons.

    The final embedding, denoted as $e_q$, is formed by combining the embeddings of the point of interest (POI) with the user, the POI with the season, and the time with the category. This composite embedding effectively represents the essential aspects of a check-in operation. Each trajectory input is comprised of a sequence of check-in embeddings and is denoted as $q = \langle p, u, t \rangle$, where the point of interest (POI) $p$ belongs to category $c$. In order to provide accurate suggestions for points of interest (POI), the transformer encoder enhances these embeddings by extracting intricate patterns and insights.

    \subsection{Transformer Encoder and MLP Decoders}\label{subsec:tran-enc-mlp-dec}
        \subsubsection{Transformer Encoder}
            The transformer encoder is a crucial component of our system. It is composed of stacked layers with positional encoding and plays a vital role in the subsequent Point of Interest (POI) recommendation process. The input tensor \(X^{[0]}\) is created by combining the historical check-in embeddings for each trajectory \(S_u\), resulting in a tensor of size \(k \times d\), where \(d\) is the embedding dimension. The utilization of normalization and residual connections is combined with fully connected networks and multi-head self-attention mechanisms in each layer. The encoder layer generates an output denoted as \(X^{[l+1]} \in \mathbb{R}^{k \times d}\) via a sequence of transformations. Several of the enhancements include:
            
            \begin{equation}
                S = X^{[l]}W_q (X^{[l]}W_k)^T \in \mathbb{R}^{k \times k}
            \end{equation}
            \textbf{Purpose:} Compute a Similarity Matrix to capture the connections between various embeddings.
            
            \begin{equation}
                S'_{i,j} = \frac{\exp(S_{i,j})}{\sum_{j=1}^{k} \exp(S_{i,j})}
            \end{equation}
            \textbf{Purpose:} Rescale the similarity values in the Similarity Matrix to make them more uniform.
            
            \begin{equation}
                \text{head}_1 = S'X^{[l]}W_v \in \mathbb{R}^{k \times \frac{d}{h}}
            \end{equation}
            \textbf{Purpose:} Calculate the Attention Output in order to determine the significant embeddings.
            
            \begin{equation}
                \text{Multihead}(X^{[l]}) = [\text{head}_1; \ldots; \text{head}_h] \times W_o \in \mathbb{R}^{k \times d}
            \end{equation}
            \textbf{Purpose:} Integrate information from multiple perspectives by combining attention heads.
            
            \begin{equation}
                X^{[l]}_{\text{attn}} = \text{LayerNorm}(X^{[l]} + \text{Multihead}(X^{[l]}))
            \end{equation}
            \textbf{Purpose:} Implement Layer Normalization to provide stable learning.
            
            \begin{equation}
                X^{[l]}_{\text{FC}} = \text{ReLU}(W_1X^{[l]}_{\text{attn}} + b_1)W_2 + b_2 \in \mathbb{R}^{k \times d}
            \end{equation}
            \textbf{Purpose:} Utilize a Feed-Forward Network to effectively capture sophisticated patterns.
            
            \begin{equation}
                X^{[l+1]} = \text{LayerNorm}(X^{[l]}_{\text{attn}} + X^{[l]}_{\text{FC}}) \in \mathbb{R}^{k \times d}
            \end{equation}
            \textbf{Purpose:} Acquire the ultimate result by consolidating data for the subsequent layer.
            
        \subsubsection{MLP Decoders}
            The multi-layer perceptron (MLP) decoders play a crucial role in predicting the next point of interest (POI), the visiting time, and the category of the POI. These decoders get input from the output of the transformer encoder. There exist three distinct MLP heads that do these predictions. The final recommendation is obtained by merging the output from the POI head with the transition attention map. To clarify, the time head is responsible for modeling the time intervals between check-ins, while the category head determines the forecasts for the subsequent points of interest (POI). Let \( X^{[l^*]} \) represent the output of the encoder. The MLP heads can be expressed as:
            
            \begin{equation}
                \hat{Y}_{\text{poi}} = X^{[l^*]}W_{\text{poi}} + b_{\text{poi}}
            \end{equation}
            \textbf{Purpose:} Utilize predictive modeling to determine the next point of interest (POI).
            
            \begin{equation}
                \hat{Y}_{\text{time}} = X^{[l^*]}W_{\text{time}} + b_{\text{time}}
            \end{equation}
            \textbf{Purpose:} Utilize the Predict Visiting Time feature to accurately estimate the time of a visit.
            
            \begin{equation}
                \hat{Y}_{\text{cat}} = X^{[l^*]}W_{\text{cat}} + b_{\text{cat}}
            \end{equation}
            \textbf{Purpose:} Predict the POI Category in order to determine the category of the next POI.

    \subsection{Operational Time Filter}\label{subsec:operational-filter}
    Following the computations of the MLP Decoders layers, a crucial step is to eliminate points of interest (POIs) that are not currently functioning. These POIs should have their selection probability reduced to zero because they are inactive at the present moment, even if they may be active at other times. The filter layer receives the output of the previous layers, denoted as \( \hat{Y} \), and applies a function \( \text{Filter}(\cdot) \) to modify the probability of selecting points of interest (POIs) accordingly. The modified probabilities, represented as \( \hat{Y}_{\text{filtered}} \), are then employed for the ultimate suggestion.
    
    The function \( \text{Filter}(\cdot) \) is defined in the following manner:
    \begin{equation}
        \hat{Y}_{\text{filtered}} = \text{Filter}(\hat{Y})
    \end{equation}
    \( \hat{Y} \) represents the initial probabilities of selecting a point of interest (POI), whereas \( \hat{Y}_{\text{filtered}} \) reflects the probabilities after applying the operational time filter.
        
    \subsection{Loss}
    The loss function, which measures prediction accuracy over several factors, is essential to model training. The loss function specifically incorporates cross entropy for both the temporal prediction and the point of interest (POI) category predictions, together with mean squared error (MSE) for temporal prediction. An amplification technique is used, in which the temporal loss is weighted by a factor of 10 in order to handle the difficulties of temporal prediction and preserve balanced gradients. While reducing the effect of other loss components, this modification gives priority to temporal changes during optimization.
        
    The final loss function can be formally expressed by the subsequent equation:
        
    \begin{equation}
        L_{\text{final}} = L_{\text{poi}} + 10 \times L_{\text{time}} + L_{\text{cat}}
    \end{equation}
        
    The variables $L_{\text{poi}}$, $L_{\text{time}}$, and $L_{\text{cat}}$ represent the individual loss contributions that arise from predicting points of interest (POI), temporal factors, and POI categories, respectively.

%
%
%

%%% Experiment setup
\section{Experimental Setup}

    \subsection{Experimental Environment}
    
    High-end settings are necessary for deep learning models to enable efficient parallel processing. Consequently, we have utilized Google Colab \cite{carneiro2018performance} for our purposes. The platform is a cloud-based Jupyter notebook that offers the essential features to utilize GPU and TPU. The Tesla T-4 GPU from NVIDIA, with 12 GB of GPU RAM, is compatible with the Ubuntu operating system. It provided the Python runtime and necessary pre-configured libraries and packages to execute deep learning tasks.

    \subsection{Optimizing hyperparameters}
    Hyperparameters have an impact on the way weights are initialized and the order in which input is processed. Therefore, identifying the most crucial values for hyperparameters enhances the accuracy of our predictive model. Table \ref{tab:hyper-param} presents the optimal settings for our classifier. The primary hyperparameters that have the largest impact on a transformer-based model include the learning rate, batch size, and number of epochs. In our proposed model, we optimize these hyperparameters: learning rate = 0.001, batch size = 16, epoch = 200, dropout = 0.3, learning rate scheduler factor = 0.1, and weight decay = 5e-4. The optimizer employed is AdamW, as proposed by Loshchilov and Hutter in their 2017 paper "Decoupled Weight Decay Regularization" \cite{loshchilov2017decoupled}.

    \begin{table}[H]
        \caption{Model Parameters and values}
        \label{tab:hyper-param}
        \centering
        \begin{tabular}{|l|l|}
            \hline
            \multicolumn{1}{|c|}{\textbf{Hyperparameters}} & \multicolumn{1}{c|}{\textbf{SEAGET}}\\ \hline
            learning\_rate (AdamW) & 1e-03 \\ \hline
            batch\_size & 16 \\ \hline
            epoch & 200 \\ \hline
            dropout & 0.3 \\ \hline
            lr\_scheduler\_rate & 0.1 \\ \hline
            weight\_decay & 5e-04 \\ \hline
        \end{tabular}
    \end{table}

    \subsection{Dataset}
    We conducted a thorough analysis of the FourSquare-NYC public dataset, which was curated by Dingqi et al. \cite{Yang2014Model}. This dataset covers the time period from April 2012 to February 2013 and provides information about various user interactions in different parts of New York City. Each dataset entry contains essential information such as user identification, visited place of interest (POI), POI categorization, GPS coordinates, and interaction timestamp.
        \subsubsection{Data Prepossessing}
        \label{sec:dp}
        To assure statistical significance, we filtered the dataset by excluding POIs and users with fewer than ten check-ins. To capture both temporal continuity and spatial diversity, we then divided the users' check-in behaviors into segments, each separated by 24 hours. To increase the analysis's precision, outliers—which were recognized as single check-ins—were eliminated.
        
        Using a single-year dataset, we randomly divided 80\% of the trajectories into training, 10\% into validation, and 10\% into testing in order to capture seasonal influences. Selecting the first 80\% of check-ins for training in a sequential manner would have resulted in seasonal bias, but our randomization made sure that all seasonal trends were represented in the training data.
        
        Finally, in order to ensure that persons or POIs not observed during training were not included in the final assessment, we used rigorous exclusion criteria during evaluation. By doing this, bias and overfitting were reduced, and the predictive performance of the model was enhanced.
        
        Important statistical information from the dataset is presented in Table~\ref{tab: dataset}.
        
        \begin{table}[h]
            \caption{Dataset Statistics}\label{tab: dataset}
            \centering
            \begin{tabular}{|l|l|l|l|l|}
                \hline
                user & poi & cat & check-in & trajectory \\
                \hline
                1,075 & 5,099 & 318 & 104,074 & 14,160\\
                \hline
            \end{tabular}
        \end{table}

%
%
%

%%% Results part
\section{Experimental Evolution}

    \subsection{Evaluation Metrics}
    We employed sophisticated metrics to assess the effectiveness of our recommendation system, specifically targeting two commonly employed indicators in recommender systems: \textit{Mean Reciprocal Rank (MRR)} and \textit{Accuracy@k (Acc@k)}. These metrics provide vital insights into the system's capacity to suggest relevant points of interest (POIs) to uers. Below is the explanation of these metrices:

        \subsubsection{Accuracy@k}
        The metric \textit{Accuracy@k} evaluates the accuracy of a system by verifying if the actual point of interest (POI) is included in the top-\(k\) recommended POIs. In a formal manner, it is computed as:

        \begin{equation}
            \text{Acc@k} = \frac{1}{m} \sum_{i=1}^{m} \mathbb{I}(\text{rank} \leq k)
        \end{equation}
        
        Here, \(m\) represents the total number of samples or trajectories, and the indicator function \(1\) determines whether the rank of the real POI is inside the top \(k\) positions.
        
        \subsubsection{Mean Reciprocal Rank (MRR)}
        \textit{Mean Reciprocal Rank (MRR)} takes into account the placement of the accurate recommendation inside the ordered list. The calculation is as follows:
        
        \begin{equation}
            \text{MRR} = \frac{1}{m} \sum_{i=1}^{m} \frac{1}{\text{rank}}
        \end{equation}
        
        The term \textit{rank} refers to the specific position of the next real Point of Interest (POI) in the sorted list. \textit{Mean Reciprocal Rank (MRR)} offers valuable insights about the system's effectiveness in ranking and prioritizing relevant Points of Interest (POIs).
        
        Increased values of \textit{Accuracy@k} and \textit{MRR} signify superior system performance, demonstrating its capacity to precisely suggest relevant POIs. By optimizing these measures, we may increase user satisfaction and engagement, hence enhancing the overall effectiveness and usefulness of the recommendation system.

\subsection{Proposed Model's Performance}
\begin{figure*}[ht!]
    \centering
    \subfloat[Top-1 accuracy for different settings of parameters.]{
        \includegraphics[width=0.48\linewidth]{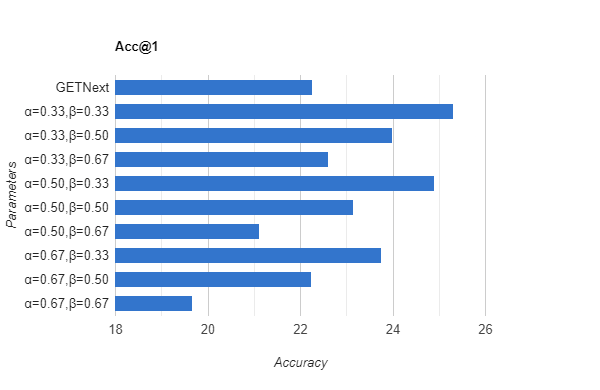}
        \label{fig:acc-1}
    }
    \hfill
    \subfloat[Top-5 accuracy for different settings of parameters.]{
        \includegraphics[width=0.48\linewidth]{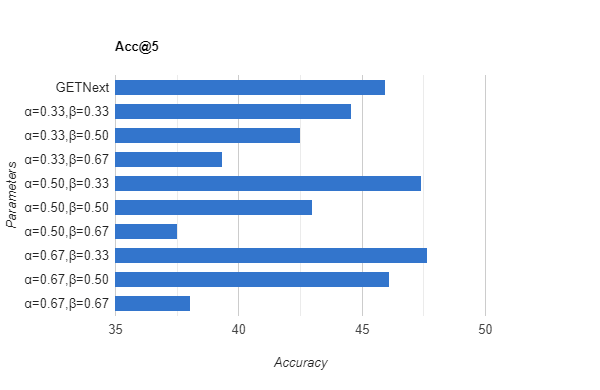}
        \label{fig:acc-5}
    }
    \vfill
    \subfloat[Top-10 accuracy for different settings of parameters.]{
        \includegraphics[width=0.48\linewidth]{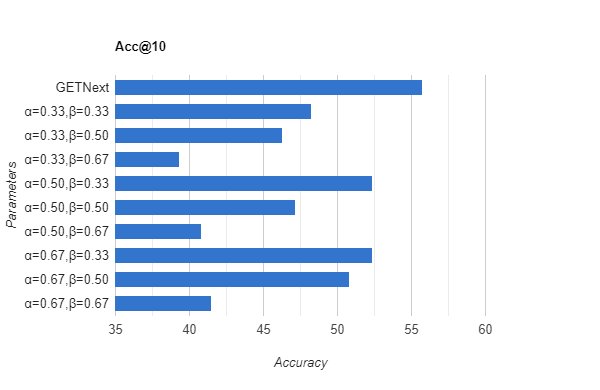}
        \label{fig:acc-10}
    }
    \hfill
    \subfloat[Top-20 accuracy for different settings of parameters.]{
        \includegraphics[width=0.48\linewidth]{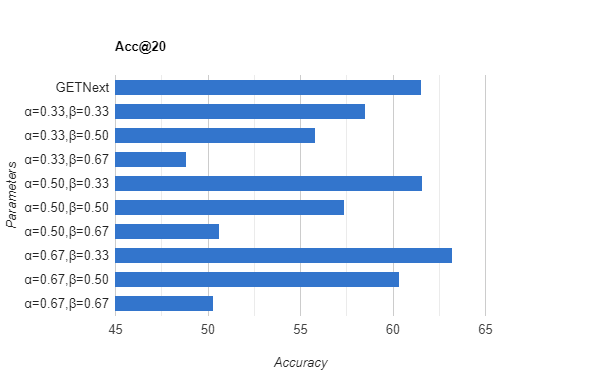}
        \label{fig:acc-20}
    }
    \caption{Bar charts of all four accuracy metrices.}
    \label{fig:all-acc}
\end{figure*}

We have conducted a number of well designed experiments that have contributed significantly to our understanding of the complex dynamics of our model and marked our journey through the dataset. Table~\ref{tab:results} presents the empirical evidence that we gathered during our extensive experimentation. The highlights of our model's performance are arranged in columns and rows. By applying a critical eye to statistical analysis, we compared the many combinations of $\alpha$ and $\beta$ and uncovered the complex dynamics between these variables. All the accuracy metrices are shown as bar charts in figure~\ref{fig:all-acc}.

\begin{table*}[ht]
    \centering
    \caption{Experimental Results}
    \begin{tabular}{|c|c|*{5}{c|}}
        % Header
        \hline
        \multicolumn{2}{|c|}{} & \textbf{Acc@1} & \textbf{Acc@5} & \textbf{Acc@10} & \textbf{Acc@20} & \textbf{MRR} \\

        % Baseline
        \hline
        \multicolumn{7}{|c|}{\textbf{Baseline Models}} \\

        \hline
        \multicolumn{2}{|c|}{\textbf{MF}} & 0.0368 & 0.0961 & 0.1522 & 0.2375 & 0.0672 \\

        \hline
        \multicolumn{2}{|c|}{\textbf{FPMC}} & 0.1003 & 0.2126 & 0.2970 & 0.3323 & 0.1701 \\

        \hline
        \multicolumn{2}{|c|}{\textbf{LSTM}} & 0.1305 & 0.2719 & 0.3283 & 0.3568 & 0.1857 \\

        \hline
        \multicolumn{2}{|c|}{\textbf{PRME}} & 0.1159 & 0.2236 & 0.3105 & 0.3643 & 0.1712 \\

        \hline
        \multicolumn{2}{|c|}{\textbf{ST-RNN}} & 0.1483 & 0.2923 & 0.3622 & 0.4502 & 0.2198 \\

        \hline
        \multicolumn{2}{|c|}{\textbf{STGN}} & 0.1716 & 0.3381 & 0.4122 & 0.5017 & 0.2598 \\

        \hline
        \multicolumn{2}{|c|}{\textbf{STGCN}} & 0.1799 & 0.3425 & 0.4279 & 0.5214 & 0.2788 \\

        \hline
        \multicolumn{2}{|c|}{\textbf{PLSPL}} & 0.1917 & 0.3678 & 0.4523 & 0.5370 & 0.2806 \\
        
        \hline
        \multicolumn{2}{|c|}{\textbf{GETNext}} & 0.2225 & 0.4593 & \textbf{0.5574} & 0.6156 & 0.3293 \\

        \hline
        \multicolumn{7}{|c|}{} \\

        % Our Model
        \hline
        \bm{$\alpha$} & \bm{$\beta$} & \multicolumn{5}{|c|}{\textbf{SEAGET}} \\

        % alpha = 0.33
        \hline
        \multirow{3}{*}{\centering\textbf{0.33}} & \textbf{0.33} & \textbf{0.2530} & 0.4457 & 0.4825 & 0.5851 & 0.3337 \\
        \cline{2-7}
         & \textbf{0.50} & 0.2398 & 0.4250 & 0.4629 & 0.5580 & 0.3157 \\
        \cline{2-7}
         & \textbf{0.67} & 0.2260 & 0.3632 & 0.3935 & 0.4882 & 0.2913 \\
        \hline
        
        % alpha = 0.50
        \hline
        \multirow{3}{*}{\centering\textbf{0.50}} & \textbf{0.33} & 0.2490 & 0.4742 & 0.5237 & 0.6157 & \textbf{0.3413} \\
        \cline{2-7}
         & \textbf{0.50} & 0.2315 & 0.4298 & 0.4713 & 0.5736 & 0.3129 \\
        \cline{2-7}
         & \textbf{0.67} & 0.2111 & 0.3753 & 0.4079 & 0.5062 & 0.2830 \\
        \hline

         % alpha = 0.67
        \hline
        \multirow{3}{*}{\centering\textbf{0.67}} & \textbf{0.33} & 0.2375 & \textbf{0.4765} & 0.5236 & \textbf{0.6319} & 0.3329 \\
        \cline{2-7}
         & \textbf{0.50} & 0.2223 & 0.4611 & 0.5084 & 0.6037 & 0.3176 \\
        \cline{2-7}
         & \textbf{0.67} & 0.1966 & 0.3803 & 0.4149 & 0.5029 & 0.2757 \\
        \hline
        
    \end{tabular}
    \label{tab:results}
\end{table*}

\begin{table*}[ht]
    \centering
    \begin{minipage}{\textwidth}
        \footnotesize\textit{Note: The performance metrics of the SEAGET model for various combinations of $\alpha$ and $\beta$ parameters are shown here. The exact values of $\alpha$ and $\beta$ for each matching row are shown in the first two columns, respectively.}
    \end{minipage}
\end{table*}

%%% bar charts

\subsection{Comparison with Existing Models}
Baseline models we utilized to compare with our model:
    \begin{itemize}
        \item MF \cite{koren2009matrix} is one of the traditional approaches for many recommendation tasks. Using matrix factorization, it discovered the latent representation of users and points of interest.
        \item FPMC \cite{Rendle2010Factor} integrated Matrix Factorization and Markov Chain to identify sequential behavior as well as long-term user preferences.
        \item LSTM \cite{Sepp1997Long} is a RNN model variation for handling sequential data. Both short- and long-term sequential patterns are modeled by LSTM in contrast to the conventional RNN model.
        \item PRME \cite{Feng2015Personalized} suggested a pairwise embedding technique called customized ranking metric embedding in order to capture user choice and the sequential transition between points of interest.
        \item ST-RNN \cite{Liu2016Pred} used the time, distance transition matrix to describe the local temporal and spatial contexts in addition to an RNN for collecting user sequential patterns.
        \item STGN \cite{Zhao2019Where} expands the traditional LSTM by incorporating temporal and spatial gates in order to capture user preferences in both space and time.
        \item STGCN \cite{Zhao2019Where} is an upgraded STGN that makes use of forget gates and coupled input.
        \item PLSPL \cite{Wu2022PLong} used LSTM to learn the user's short-term preference and attention mechanism to learn their long-term preferences, which were then integrated by personalized linear layers.
        \item GETNext \cite{Yang2022GETNext} makes use of a Graph Enhanced Transformer model (GETNext) and a user-agnostic global trajectory flow map to better leverage the vast collaborative signals for a more precise next POI prediction.
    \end{itemize}

For different combinations of \(\alpha\) and \(\beta\), our model outperforms the baseline models consistently on all sorts of accuracy tests except the \textbf{Acc-10}. This exhaustive assessment proves our method's exceptional usefulness and effectiveness.

The robustness and flexibility of our model in capturing the complex dynamics of user behavior and POI popularity is particularly shown by its capacity to adjust to varying values of \(\alpha\) and \(\beta\). The findings show that not only does our model perform better when it comes to predicting the next point of interest (POI), but it also stays accurate across a variety of circumstances and parameter configurations.

The improved accuracy metrics highlight our model's major advancements and demonstrate its capacity to deliver recommendations that are more timely and relevant. This development is critical for practical applications because the temporal and relevancy of recommendations can have a significant influence on user engagement and satisfaction.

In general, our model's higher performance in accuracy tests compared to the baseline models confirms its potential as a front-runner in the POI recommendation systems domain.

%
%
%

%%% Discussion part
\section{Discussion}
In today's entertainment and travel industries, Point of Interest (POI) recommendation systems are increasingly crucial. We have attempted to provide insight into the application of an advanced POI recommendation system in this research.

Our study has shown how important it is to consider seasonal variations, and POI operating hours when choosing which POI to visit next. Our SEAGET model performs much better than the state-of-the-art techniques, as shown by the findings in Table \ref{tab:results}, demonstrating the value of including these factors.

Numerous obstacles were faced throughout the development of the SEAGET model. One major constraint was the lack of a suitable dataset that included all the required data for our research. The dataset utilized, however the most appropriate option, comprises data from solely one year. Having data that covers numerous consecutive years would have facilitated a more prominent evaluation of seasonal influences. Furthermore, the dataset does not include information regarding the specific hours of operation for points of interest (POIs), which requires the human gathering of general operational hours for different kinds of POIs in the cities that were analyzed. Adding real-time operational hours data would undoubtedly improve the model's performance.

There are particular limitations to our model. For example, the patterns of check-ins usually differ on days when there is less activity, but our model treats all days in the same way, and as a result, it fails to accurately represent these fluctuations. User visitor patterns may deviate from typical trends and might be influenced by a multitude of events and conditions. The restricted dataset presented a barrier in capturing these dynamic changes, which we were unable to fully address. In addition, certain Points of Interest (POIs) may be momentarily unreachable due to variables such as road construction, which our current model is unable to handle.

Another limitation is the possibility of congestion at popular Points of Interest (POIs), while others continue to be neglected. Our current strategy is not efficiently achieving a balanced distribution of customers across multiple Points of Interest (POIs) within the same category.

Despite these difficulties and restrictions, the progress demonstrated in our SEAGET model highlights the possibility for additional investigation and improvement in this field. Future research should prioritize overcoming these limitations by integrating more extensive datasets and devising methods to dynamically adjust recommendations in response to real-time fluctuations and user-specific subtleties. This has the potential to result in more precise and user-focused systems for recommending points of interest (POI).

%
%
%

%%% Conclusion part 
\section{Conclusion and Future Work} 
The objective of this study was to emphasize the important significance of seasonal variations in the selection of points of interest (POI), illustrating that seasonality is a critical element in deciding the subsequent destination to visit. The SEAGET model we propose explores a novel research direction that specifically examines the effects of seasonal factors on recommendations for points of interest (POI).

One significant contribution of our work is the implementation of an operating hour-based filtering strategy. This technique assures that the recommended points of interest (POIs) are open and available for visitation during the specified period, hence improving the overall performance of the model.

We also revised the notion of popularity attributed to points of interest (POIs) in order to make advancements in the field of POI recommendation systems. Through a rigorous analysis of our model's dynamics using a comprehensive real-world dataset, we examined the detailed relationship between the model's performance and the parameters $\alpha$ and $\beta$ defined in Equation \ref{eq:pop} over multiple $\alpha$ and $\beta$ values. This meticulous experimentation yielded vital insights into the intricate behavior of the recommendation system.

Our research signifies the commencement of a more extensive endeavor to reveal the diverse aspects of POI recommendations. Subsequent research will focus on further defining the precise concept of popularity. In addition, we will integrate commercial perspectives into the model, with the goal of evenly distributing check-ins among similar category points of interest (POIs). In order to improve the accuracy of our recommendations and ensure customer happiness, we will also take into account other contextual aspects such as weekdays, weekends, and festivals.

%
%
%

%%% Acknowledgement part 
\section{Acknowledgments}
The author gratefully acknowledges the financial support from the Ministry of Science and Technology, Bangladesh with the National Science and Technology (NST) Fellowship, which has contributed to the research presented in this paper.

%
%
%

%%% Bibliography part
%% Loading bibliography style file
%\bibliographystyle{model1-num-names}
%\bibliographystyle{model5-names}
\bibliographystyle{cas-model2-names}
%\bibliographystyle{IEEEtran}

%% Loading bibliography database
\bibliography{cas-refs.bib}

\end{document}